\documentclass[
reprint,
superscriptaddress,
 amsmath,amssymb,
 aps,
]{revtex4-2}

\usepackage{graphicx}% Include figure files
\usepackage{dcolumn}% Align table columns on decimal point
\usepackage{bm}% bold math
\usepackage{tabularx}
\usepackage{hyperref}% add hypertext capabilities
\usepackage{float} % for forcing figure placements
\usepackage{comment} % 
\usepackage{xcolor}

\makeatletter
\def\maketitle{
\@author@finish
\title@column\titleblock@produce
\suppressfloats[t]}
\makeatother

\begin{document}

\preprint{APS/123-QED}

\title{Tunable anharmonicity in Sn-InAs nanowire transmons beyond the short junction limit
}
\author{Amrita Purkayastha}
\altaffiliation[Current address: ]{Materials Research Laboratory, University of Illinois at Urbana-Champaign, Urbana, IL 61801}
\affiliation{Department of Physics and Astronomy, University of Pittsburgh, Pittsburgh, PA 15260, USA}

\author{Amritesh Sharma}
\affiliation{Department of Physics and Astronomy, University of Pittsburgh, Pittsburgh, PA 15260, USA}

\author{Param J. Patel}
\affiliation{Department of Physics and Astronomy, University of Pittsburgh, Pittsburgh, PA 15260, USA}
\affiliation{Department of Applied Physics and Yale Quantum Institute, Yale University, New Haven, CT 06511, USA}

\author{An-Hsi Chen}
\affiliation{Univ. Grenoble Alpes, Grenoble INP, CNRS, Institut N\'eel, 38000 Grenoble, France}

\author{Connor P. Dempsey}
\affiliation{Department of Electrical and Computer Engineering,
University of California, Santa Barbara, CA 93106}

\author{Shreyas Asodekar}
\affiliation{Department of Physics and Astronomy, University of Pittsburgh, Pittsburgh, PA 15260, USA}

\author{Subhayan Sinha}
\affiliation{Department of Physics and Astronomy, University of Pittsburgh, Pittsburgh, PA 15260, USA}

\author{Maxime Tomasian}
\altaffiliation[Current address: ]{Univ. Grenoble Alpes, Grenoble INP, CNRS, Institut N\'eel, 38000 Grenoble, France}
\affiliation{Department of Physics and Astronomy, University of Pittsburgh, Pittsburgh, PA 15260, USA}

\author{Mihir Pendharkar}
\altaffiliation[Current address: ]{Department of Materials Science and
Engineering, Stanford University, Stanford, CA 94305}
\affiliation{Department of Electrical and Computer Engineering,
University of California, Santa Barbara, CA 93106}

\author{Christopher J. Palmstr{\o}m}
\affiliation{Department of Electrical and Computer Engineering,
University of California, Santa Barbara, CA 93106}
\affiliation{California NanoSystems Institute, University of California Santa Barbara, Santa Barbara, CA 93106, USA}
\affiliation{Materials Department, University of California Santa Barbara, Santa Barbara, CA 93106, USA}

\author{Moïra Hocevar}
\affiliation{Univ. Grenoble Alpes, Grenoble INP, CNRS, Institut N\'eel, 38000 Grenoble, France}%

\author{Kun Zuo}
\affiliation{School of Physics, The University of Sydney, Sydney, NSW 2006, Australia}
\affiliation{ARC Centre of Excellence for Engineered Quantum Systems, School of Physics,
The University of Sydney, Sydney, NSW 2006, Australia}

\author{Michael Hatridge}
\affiliation{Department of Applied Physics and Yale Quantum Institute, Yale University, New Haven, CT 06511, USA}

\author{Sergey M. Frolov$^\ast$}
\email{frolovsm@pitt.edu}
\affiliation{Department of Physics and Astronomy, University of Pittsburgh, Pittsburgh, PA 15260, USA}

\date{\today}

\begin{abstract}

The anharmonicity of a transmon qubit, defined as the difference in energy level spacing, is a key design parameter. In transmons built from hybrid superconductor-semiconductor Josephson elements, the anharmonicity is tunable with gate voltages that control both the Josephson energy and the weak link transparency. In Sn-InAs nanowire transmons, we use two-tone microwave spectroscopy to extract anharmonicity ranging in absolute value from the transmon charging energy $E_c$ to values smaller than $E_c/10$. This behavior contrasts with the predictions of the multi-channel short-junction model, which sets a lower limit on anharmonicity at $E_c/4$. Coherent operation of the qubit is still possible at the point of the lowest anharmonicity. These findings demonstrate the potential of quantum circuits that benefit from widely electrically tunable anharmonicity.

\end{abstract}

\maketitle

%\tableofcontents
\section{\label{sec:level1} Introduction}
Transmon qubits are superconductor circuits with nonlinear Josephson elements, that exhibit quantized behavior at cryogenic temperatures~\cite{devoret2013superconducting,krantz2019quantum}. Multiple quantum computing projects rely on aluminum/aluminum-oxide tunnel junctions that are either fixed-frequency or flux-tunable ~\cite{bland20252dtransmonslifetimescoherence, bal2024systematic, google2025QEC}. Alternative sources of nonlinearity are also explored, among them gate-voltage-tunable superconductor-semiconductor weak links which allow for the integration of a broader array of superconductor materials, and may deliver faster control that is more localized within the circuit~\cite{larsen2015semiconducting,Casparis_2016,Casparis2018,Luthi2018,Wang2019,sagi2024gate,Purkayastha2026SnTransmon, Aparicio_2025_gatetunable_graphene, zheng2024coherent, Zhuo2023, Kiyooka2025NanoLett_Gatemon, Gatemonium_2025, Riechert2025CNTGatemon, Strickland2024}. 

In aluminum oxide junctions, supercurrents are mediated by Cooper pair tunneling, which is well approximated by the sinusoidal Josephson relation~\cite{Josephson1962effects}. This makes it possible to design the qubit Hamiltonian using the Josephson energy $E_J$ controlled by the junction area ~\cite{AB_relation1963}, and the charging energy $E_c=e^2/2C$, where $C$ is the large capacitance shunting the junction~\cite{koch2007charge}. Of relevance to this work is the anharmonicity $ \alpha = (E_{12} - E_{01})$, which compares the energy separation of the second excited state to the qubit level splitting. The anharmonicity affects pulse speed and quantum gate fidelity \cite{koch2007charge,Motzoi2009,gambetta2006qubit}.  Fig.~\ref{fig:1}(a) shows the transmon potential compared to that of a harmonic oscillator, e.g., a $LC$ resonator. In contrast to the harmonic case, the transmon energy levels are not equally spaced, which makes it possible to selectively address the qubit states $|0\rangle$ and $|1\rangle$ with a microwave tone at frequency $f_{01}$. For tunnel-junction transmons operated in the large $E_J/E_C$ regime, the anharmonicity is approximately $\alpha \approx -E_C$~\cite{koch2007charge}.

Supercurrent in semiconductors is not due to tunneling: it is mediated by mobile charges through the proximity effect~\cite{DeGennes1964} which can be described in terms of Andreev bound states carrying supercurrent~\cite{andreev1964thermal,golubov2004current}.  An expression frequently used to calculate the weak link Josephson energy is based on adding currents through a finite number of conducting channels~\cite{Beenakar1991}:
\begin{equation}
    V(\phi)=-\Delta\sum_i\sqrt{1-\tau_i \sin^2(\phi/2)},
    \label{Eq.1}
\end{equation}
where $V$ is the Josephson potential, $\Delta$ is the induced superconducting gap, $\phi$ is the phase difference across the junction and $\tau_i$ are channel transparencies. Using this energy-phase relation Eq.~\eqref{Eq.1}, 
Ref.~\cite{KringAnharmonicity2018} showed the anharmonicity to be:
\begin{equation}
     \alpha\approx -E_c \left(1-\frac{3\sum_i \tau_i^2}{4\sum_i \tau_i}\right).
         \label{Eq.2}
 \end{equation}
 
The more ballistic and transparent junctions are characterized by energy-phase relations with higher harmonics~\cite{goldobin2007}, and this generally corresponds to a reduced nonlinearity (Fig.~\ref{fig:1}(b)). According to Eq.~\eqref{Eq.2}, the anharmonicity is minimal at $\alpha=E_c/4$ when all transparencies are set to $\tau_i=1$~\cite{KringAnharmonicity2018}. In line with this, a number of experiments have reported anharmonicites in aluminum-semiconductor weak link transmons bounded by $\alpha=E_c/4$~\cite{Hertel2022,sagi2024gate,zheng2024coherent, Liu_Bordoloi_Issokson_Levy_Vavilov_Shabani_Manucharyan_2025}.

However, Eq.~\eqref{Eq.1} is derived within the so-called short junction limit~\cite{BeenakarShort,Beenakar1991}, which is when the junction length $L$ is shorter than the induced coherence length $\xi$. This approach is an approximation because it does not include dephasing and interaction effects~\cite{likharev1979superconducting}, or weak links with quantum dots. The current-phase relation of a weak link in the long junction limit can approach a saw-tooth function (Fig.~\ref{fig:1}(b)), which drives the anharmonicity down to zero~\cite{fatemi2024nonlinearitytransparentsnsweak}. While typical superconductor-semiconductor nanowire junctions may not reach this limit, it is an open question whether Josephson coupling in them is fully described by Eq.~\eqref{Eq.1} as a sum of supercurrent-carrying channels of various transmissions.

In this work, we demonstrate that anharmonicity in Sn-InAs nanowire transmon qubits is gate-voltage tunable over a wide range from values approaching the tunnel junction limit of designed $E_c$, down to $<E_c/10$, well below the short junction transmon limit. At the point of the lowest anharmonicity $\alpha/h= 30$~MHz, we demonstrate coherent operation of the transmon with coherence times $T_1=3.3~\mu s$ and $T_{2E}=2.9~\mu s$. The anharmonicity is a smooth non-monotonic function of gate voltage. In one device we observe a close correspondence between the qubit frequency $f_{01}$ and anharmonicity, while in another device there is no such correspondence. This illustrates that anharmonicity provides additional insights into the mesoscopic current-phase relationship of nanowire weak links beyond gate-tunable qubit frequency.
These findings can lead to the development of more complete models of weak link transmon qubits and offer a path to optimizing their performance. We examine our results in the light of a recent theory and consider factors that can facilitate lower anharmonicity such as the larger superconducting gap of Sn \cite{pendharkar2021parity,sharma2025sninasnanowireshadowdefinedjosephson} compared with traditional Al \cite{Krogstrup2015Epitaxy}. Possible advantages of such gate tunable low anharmonicity could be in implementing parametric amplifiers and Kerr cat qubits \cite{grimm2020kerrcat, mirrahimi2014catqubits}. The dynamical control of qubit anharmonicity and tunable harmonic content could also be used for new two-qubit coupling schemes \cite{Yan_Krantz_Sung_Kjaergaard_Campbell_Orlando_Gustavsson_Oliver_2018, Mundada_Zhang_Hazard_Houck_2019, Ku_Xu_Brink_McKay_Hertzberg_Ansari_Plourde_2020}.  

\section{Device and Measurement Description}

\begin{figure}[!h]  
  \centering
  \includegraphics[width=\linewidth]{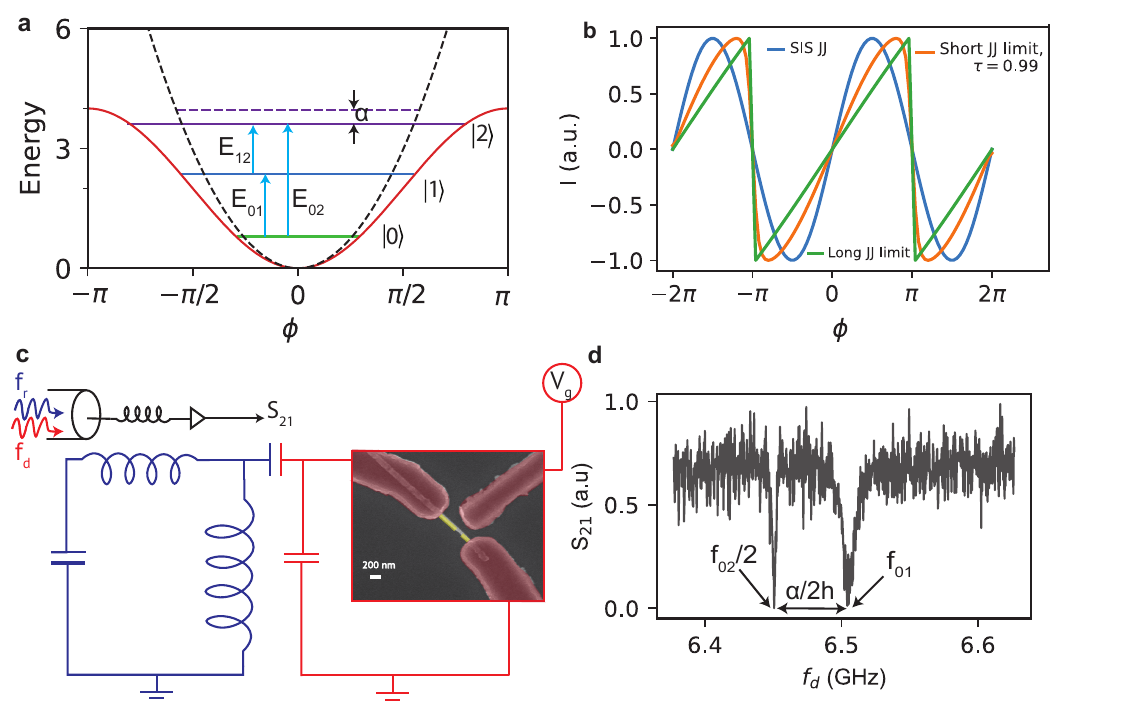}
  \caption{(a) Simplified energy diagram of a transmon (red). The black dashed curve is the parabolic potential of a harmonic oscillator. (b) Schematic current-phase relations of different weak link junctions. Insulator based junction (SIS JJ, blue), Semiconducting weak link in the short junction limit (Short JJ limit, red,  $\tau = 0.99$), long junction (Long JJ limit, green). (c) Schematic of the capacitively coupled readout resonator and transmon, overlaid with an SEM image of the nanowire junction and aluminum leads with side gate (false color). (d) Transmission in two tone spectroscopy with transitions and anharmonicity indicated.}
  \label{fig:1}
\end{figure}

The schematic of the device is presented in Fig.~\ref{fig:1}(c). It consists of an Sn-InAs nanowire Josephson junction shunted by a capacitor in the transmon geometry. The Josephson junction is made etch-free using an \textit{in-situ} shadowing technique described in Ref.~\cite{sharma2025sninasnanowireshadowdefinedjosephson}. 
The junction length is $\text{70-120 nm}$.
The charging energy for the transmon is designed to be $E_c/h \approx 380$ MHz, calculated using ANSYS HFSS using silicon permittivity $\epsilon=\text{11.7}$ (see supplementary information). The qubits are capacitively coupled to $\lambda/4$ coplanar waveguide resonators with $Q \sim 2000$ and bare resonator frequency of $f_r=8.1780$ GHz for Device A (presented in the main text) and $f_r=7.7003$ GHz for Device B (presented in supplementary). The readout resonators are inductively coupled to a common feedline enabling multiplexed dispersive readout. The capacitor, the resonator and the feedline are etched out of a 120~nm of NbTiN film on a high resistivity silicon wafer. The NbTiN groundplane is connected to the nanowire with an Al liftoff patch for devices A and B, and by an NbTiN lift-off patch for devices E and F shown in the supplementary information.

\begin{figure*}[!ht]  
  \centering
  \includegraphics[width=15cm]{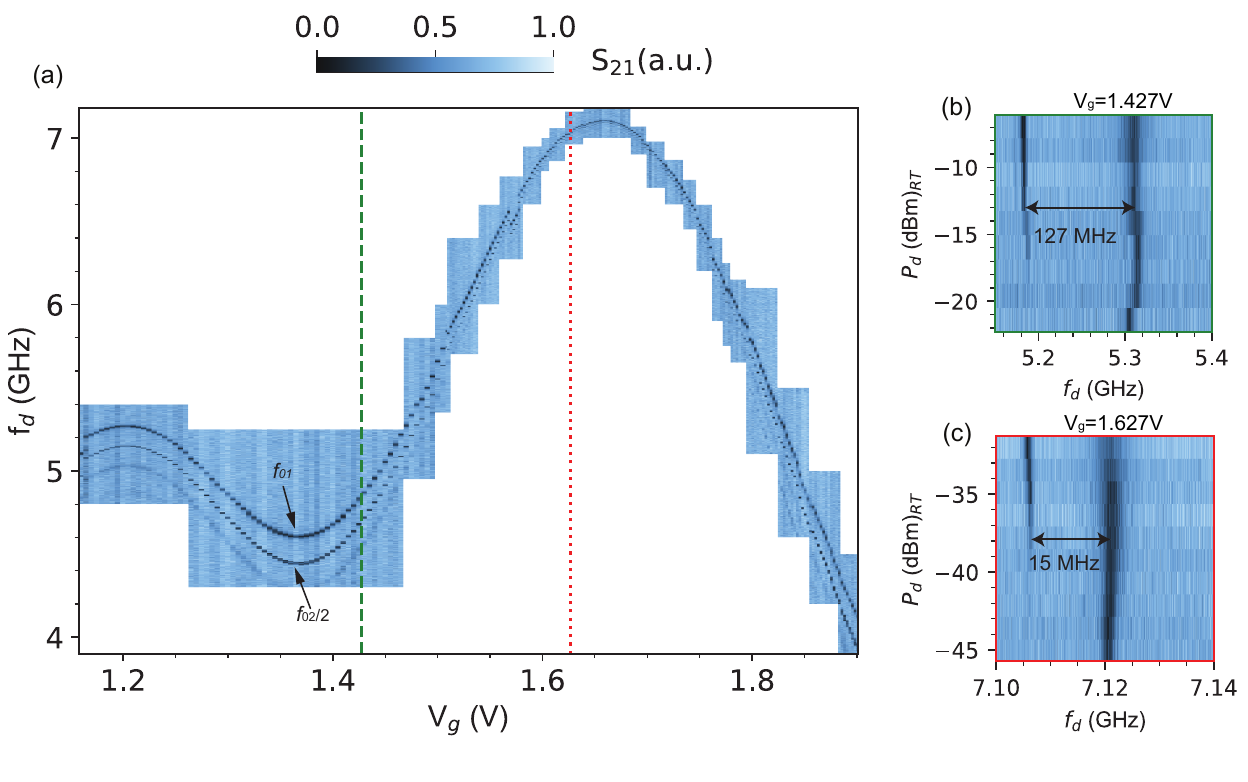}
  \caption{Device A: (a) Two-tone spectroscopy as a function of side gate voltage $V_g$ swept from low to high. Both $f_{01}$ and $f_{02}/2$ transitions are visible and changing non monotonically with $V_g$. 
   (b) and (c) Two-tone spectroscopy as a function of drive power and drive frequency at $V_g=1.427V$ and $V_g=1.627V$ respectively showing the clearly resolvable $f_{01}$ and $f_{02}/2$ dips. The separation between $f_{01}$ and $f_{02}/2$ for (b) and (c) are 127 MHz and 15 MHz respectively.}
  \label{fig:2}
\end{figure*}

Standard microwave measurements are used for characterizing the qubits, including the two-tone spectroscopy. The coherence times of  qubits A and B are reported in Ref.~\cite{Purkayastha2026SnTransmon} including $T_1 \approx 27 \mu s$ for Qubit A. At each $V_g$, the readout resonator frequency $f_r$ is first determined using single-tone spectroscopy. The transmission at $f_r$ was then monitored using a vector network analyzer (VNA) while a second tone with frequency $f_d$ was applied with a microwave generator.  Fig.~\ref{fig:1}(c) shows the measurement circuit schematic. When $f_d$ matches $f_{01}$ a qubit-state-dependent shift in $f_r$ is recorded. At lower drive power, the qubit transition from $|0\rangle$ to $|1\rangle$ appears as a dip/peak in transmission phase/amplitude at $f_{01}$. At higher drive powers, the two-photon transition is visible at a frequency $f_{02}/2$.  From the difference between $f_{02}/2$ and $f_{01}$ we extract the qubit anharmonicity as follows. We first note that:
    $f_{02} = f_{01} + f_{12}$ and 
    $\alpha/h = f_{12} - f_{01}$.
Combining the two equations,
   $ \alpha = 2h \left(\frac{f_{02}}{2} - f_{01}\right)$.
Fig.~\ref{fig:1}(d) shows transmission data with sufficiently high power to identify two of the highest peaks as the 0-1 transition at $f_{01}$ and the two-photon 0-2 transition at $f_{02}/2$.

\section{Results: Gate-dependent anharmonicity}

\begin{figure}

  \includegraphics[width=\linewidth]{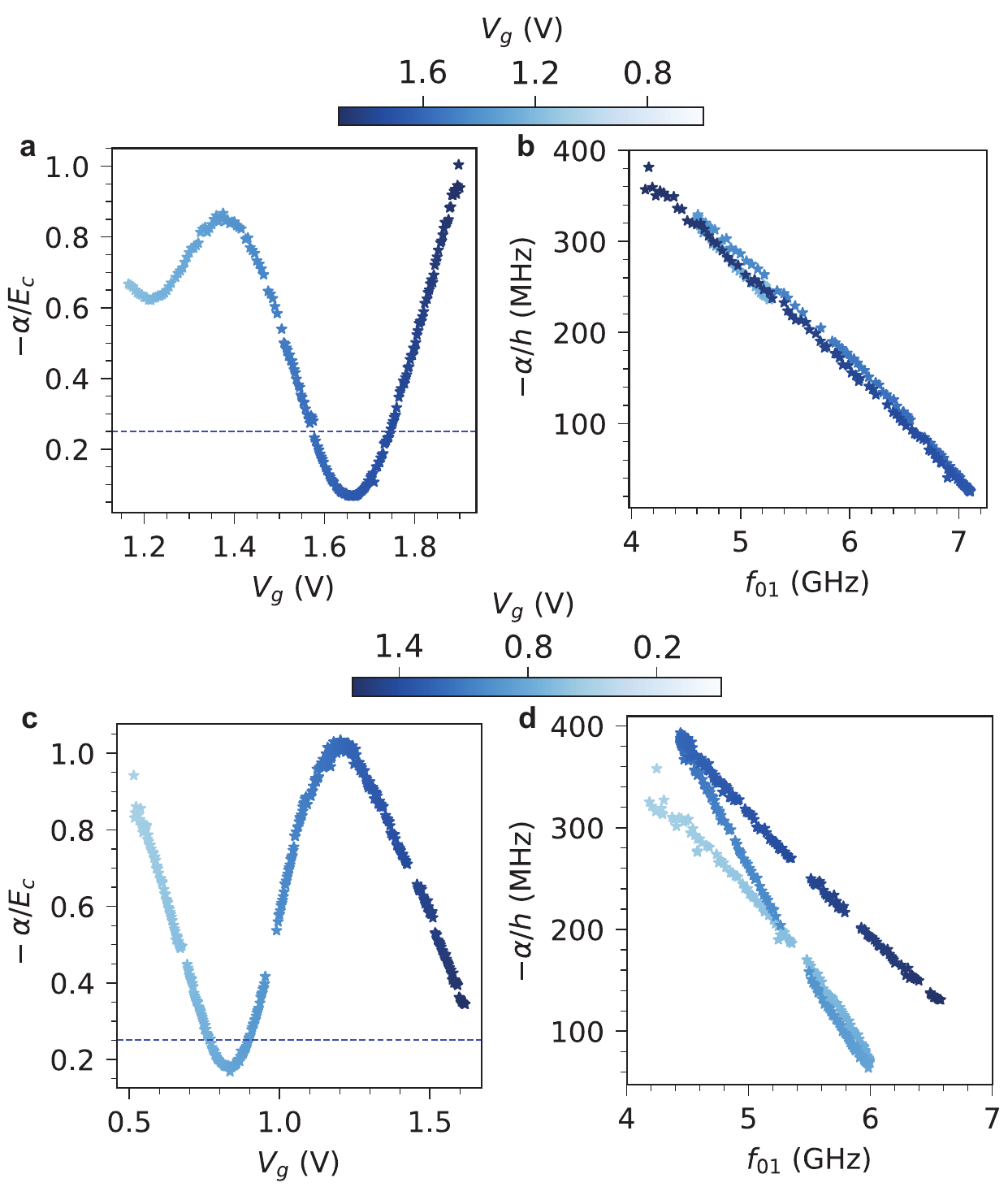}
  \caption{(a) Anharmonicity normalized with the designed charging energy, $\alpha/E_c$ plotted as a function of gate voltage $V_g$, for different qubit frequencies $f_{01}$ indicated by the color bar.(b)Anharmonicity $\alpha/h$ as a function of $f_{01}$, with corresponding gate voltages indicated by the color bar, for Device A.(c) and (d) show the same plots as in (a) and (b), respectively, for Device B.}
  \label{fig:3}
\end{figure}

Fig.~\ref{fig:2}(a) shows the map of two-tone spectroscopy for Qubit A obtained at higher drive power. The 0-1 transition at the qubit frequency $f_{01}$ and the two-photon transition at $f_{02}/2$ are changing smoothly and non-monotonically as a function of the side gate voltage $V_g$. The third resonance visible below the $f_{02}/2$ line is well described by $f_{02}-f_{01}$. In Figs.~\ref{fig:2}(b) and \ref{fig:2}(c), the dips in transmission at $f_{01}$ and $f_{02}/2$ can be resolved with increasing drive power. These panels illustrate the gate tunability of $\alpha$. The lowest observed $\alpha/h = 30$ MHz is less than $0.1E_C$ (Fig.~\ref{fig:2}(c)).

Frequencies $f_{01}$ and $f_{02}/2$ are extracted by fitting two independent Lorentzian line shapes to the two-tone signal, and their difference is $\alpha/2h$. Fig.~\ref{fig:3}(a) shows the anharmonicity $\alpha/h$   plotted as a function of $V_g$ for Qubit A. For the gate range $1.62~\text{V} < V_g < 1.7~\text{V}$, the normalized anharmonicity, $\alpha/E_c$, drops below 0.25. Fig.~\ref{fig:3}(b) shows the same data as in Fig.~\ref{fig:3}(a) but with $\alpha/h$ versus $f_{01}$ extracted from Fig.~\ref{fig:2}(a). In this case, the $\alpha/h$ values retrace, meaning that there is a unique anharmonicity for each qubit frequency. 

For Qubit B, the anharmonicity also drops below the minimum short junction value over a range of gate voltages (Fig.~\ref{fig:3}(c) and supplementary information). However, there is no one-to-one correspondence between the qubit frequency and anharnomicity: while $\alpha$ evolves smoothly over monotonic segments in $\alpha(V_g)$, between the segments it switches to a different trend line creating a zigzag pattern (Fig.~\ref{fig:3}(d)).

\begin{figure}
  \includegraphics[width=\linewidth]{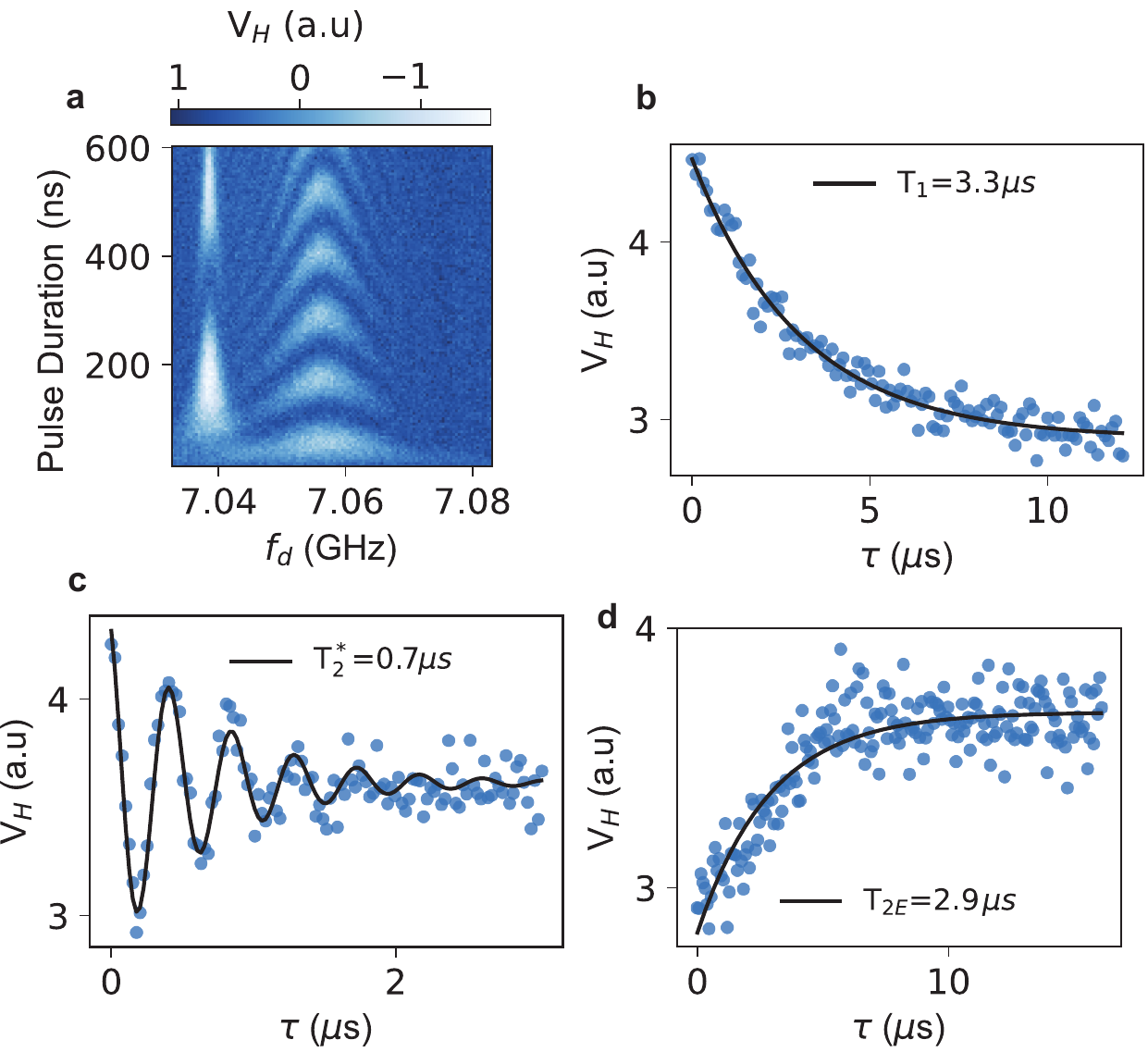}
  \caption{(a)Rabi oscillations as function of pulse duration and drive frequency at fixed drive amplitude of 0.64 a.u. (b) Relaxation time $T_1$ extracted from fit to decaying excited state population. (c) Ramsey measurements and fit to a decaying oscillation. (d) The Hahn echo sequence and fit to a decaying function. Demodulated transmission amplitude is given in arbitrary units (a.u.).
}
  \label{fig:4}
\end{figure}

We also perform time domain measurements at the qubit frequencies where the lowest anharmonicity is observed. For Qubit A, within the gate voltage range \( 1.62~\text{V} < V_g < 1.69~\text{V} \), the qubit frequency is in the range \( f_{01} = 7.05\text{-}7.15~\text{GHz} \). In Fig.~\ref{fig:4}(a) we demonstrate that both the primary qubit transition, as well as the two-photon transition can be driven coherently. Qubit coherence is measured at the lowest anharmonicity point, $f_{01} = 7.15\ \text{GHz}$, where $\alpha/h \approx 30\ \text{MHz}$ as shown in Fig.~\ref{fig:2}(c). 

The relaxation time $T_1$ is obtained by fitting the decay of the excited state population as a function of measurement delay $\tau$, yielding $T_1 = 3.3\ \mu\text{s}$ (Fig.~\ref{fig:4}(b)). Ramsey sequence measurements exhibit coherent free-induction decay oscillations, from which the dephasing time is extracted to be $T_2^* = 0.7\ \mu\text{s}$ (Fig.~\ref{fig:4}(c)). The Hahn echo sequence further extends the coherence time, with the exponential fit yielding $T_{2\mathrm{E}} = 2.9\ \mu\text{s}$ (Fig.~\ref{fig:4}(d)). 

\section{Discussion}

Generally, we observe that lower anharmonicity corresponds to higher qubit frequency. These in turn correspond to higher Josephson energies. In superconductor-semiconductor junctions this likely indicates higher channel transparencies and more transmitting channels. Even according to the short junction Eq.~\eqref{Eq.1}, higher transparency makes the energy-phase relation skewed due to the presence of higher Josephson harmonics, and hence the junctions are less nonlinear. Conversely, at lower Josephson energies the higher harmonics are suppressed and the energy-phase relation is closer to the sinusoidal behavior that is also characteristic of tunnel junctions.

Figs.~\ref{fig:3}(b) and~\ref{fig:3}(d) provide indirect evidence that the mesoscopic details of the junction affect the anharmonicity. Although in these devices we previously found a reasonable correspondence between the qubit frequency and the measured relaxation times $T_1$~\cite{Purkayastha2026SnTransmon}, the example of Qubit B shows that the anharmonicity may be sensitive to gate voltage, which controls the mode occupation and transmission, including the electron dwell time in the nanowire.

Since the anharmonicity of similar devices was previously reported within the bounds of Eq.\eqref{Eq.1}~\cite{Hertel2022,sagi2024gate,zheng2024coherent, Liu_Bordoloi_Issokson_Levy_Vavilov_Shabani_Manucharyan_2025}, we consider the question of what may be different about our junctions, such as the superconducting gap $\Delta$ of Sn which is three times larger than that of Al used in previous studies. An estimate based on junction coherence formula, $\xi = \sqrt{\xi_0 l_e}$, where $\xi_0 = \hbar v_F / \pi \Delta$ with $v_F$ the Fermi velocity, and $l_e$ the mean free path would suggest a shorter coherence length $\xi$ due to larger gap, and this can shift the junction further from the short junction limit. Note that the length of the shadow junction likely underestimates the length of the junction, since the quasiballistic nanowire segments underneath the Sn shell are also conducting. Thus, even with Al as a superconductor the junctions may not be in the short junction limit.

Anharmonicity below the minimal value allowed by the short junction formula is observed in a limited range of gate voltages. That does not mean that Eq.~\eqref{Eq.1} applies outside of that voltage range. Instead, the observations likely indicate that these junctions are not well described by Eq.~\eqref{Eq.1} and a different model is required for more complete understanding.

Motivated by our preliminary results, recent theoretical work Ref. ~\cite{fatemi2024nonlinearitytransparentsnsweak} has considered two models that go beyond Eq.~\eqref{Eq.1}, namely a short resonant level model and a ballistic finite length model. Ref.~\cite{fatemi2024nonlinearitytransparentsnsweak} calculates lower anharmonicities than the short junction model of Eq.~\eqref{Eq.1} assuming single channel Josephson weak links. From the short resonant level model the lowest calculated anharmonicity is $0.13E_c$. However, the ballistic finite length model can yield the anharmonicity as low as 0 in the limit of large junction lengths. Both models presented in Ref.~\cite{fatemi2024nonlinearitytransparentsnsweak} highlight the importance of the dwell time inside the weak link, and the interactions between the above-gap continuum energy levels and the Andreev Bound States. Other than through comparing the anharmonicity values, we do not identify a way to relate the results of these calculations to our transmon measurements, but this may evolve with future work.

In conclusion, we observe gate-dependent transmon anharmonicity that deviates significantly from the short junction model, with anharmonicity reaching values below the predicted minimum of $E_c/4$.
This indicates that qubit devices based on unconventional Josephson elements present an interesting topic of investigation in mesoscopic physics. These and future findings may bring about new quantum circuits with additional capabilities, such as single nanowire parametric amplifiers, two-qubit coupling schemes based on gate modulation of the anharmonicity \cite{Yan_Krantz_Sung_Kjaergaard_Campbell_Orlando_Gustavsson_Oliver_2018, Mundada_Zhang_Hazard_Houck_2019, Ku_Xu_Brink_McKay_Hertzberg_Ansari_Plourde_2020, Butseraen_2022}. It may also result in more compact transmons since with semiconductor weak links lower anharmonicites do not require larger capacitors.

\section{Duration and Volume of Study} 

Qubit A and Qubit B were measured over a period of four months, during which two cooldowns were performed. Qubits E and F were measured over a period of 3 months which involved two different cooldowns. In total, 22 Sn-InAs nanowire transmon devices were measured over the course of 3 years. While a consistently lower anharmonicity than designed was observed across these devices, the gate tunability of anharmonicity was not studied in detail for most of these devices. 

\section{Data availability} 
Data and code extending beyond what is presented in the main text and the supplementary are available at~\href{https://doi.org/10.5281/zenodo.16732149}{10.5281/zenodo.16732149} for devices A and B and~\href{https://doi.org/10.5281/zenodo.18715850}{10.5281/zenodo.18715850} for devices E and F.

\section{Acknowledgments} 

Nanowire growth was supported by ANR HYBRID (ANR-17-PIRE-0001), ANR IMAGIQUE (ANR-42-PRC-0047), IRP HYNATOQ and the Transatlantic Research Partnership. Sn shell growth was supported by the NSF Quantum Foundry at UCSB funded via the Q-AMASE-i program under award DMR-1906325. Nanowire microscopy characterization was supported by the U.S. Department of Energy Office of Basic Energy Sciences (BES) through grant DE-SC-0019274, and transport characterization through the U.S. Department of Energy, Basic Energy Sciences grant DE-SC-0022073. Microwave measurements were supported by the LPS/ARO nextNEQST program W911NF2210036. This work made use of the Nanoscale fabrication and Characterization facility (NFCF) at The Gertrude E. and John M. Petersen Institute of NanoScience and Engineering (PINSE) as well as of facilities at the Western Pennsylvania Quantum Information Core (WPQIC) at the University of Pittsburgh.

\section{Competing Interests}
Michael Hatridge serves as a consultant for D-Wave Inc. (formerly Quantum Circuits, Inc.), receiving remuneration in the form of consulting fees, and hold equity in the form of stock options.

\bibliography{references}
\clearpage

\clearpage

\renewcommand{\appendixname}{Supplementary Material}
\renewcommand{\thefigure}{S\arabic{figure}} \setcounter{figure}{0}
\renewcommand{\thetable}{S\arabic{table}} \setcounter{table}{0}
\renewcommand{\theequation}{S\arabic{table}} \setcounter{equation}{0}
\renewcommand{\thesection}{S\arabic{section}} \setcounter{section}{0}

\title{Supporting Information and Additional Data for "Tunable anharmonicity in Sn-InAs nanowire transmons beyond the short junction limit"}% Force line breaks with \\

\maketitle

\onecolumngrid

% \begin{figure*}[!ht]
%   \includegraphics[width=9 cm]{Figures/timerabi-low anharmonicity.pdf}
%   \caption{}
%   \label{S1-1}
% \end{figure*}

\section{Methods}

The fabrication process starts with solvent cleaning, followed by acid (Piranha+BOE) cleaning of high resistivity ($\rho > 20 k\Omega -cm$) Si-100 wafer. 
% The wafer is carried to the AJA International Inc. magnetron sputtering system in a vacuum container and loaded in the loadlock within 10 min of acid cleaning. 
A 120 nm NbTiN film is sputtered onto the wafer at $23 ^{\circ} C$ and then annealed at $600 ^{\circ} C$ for 30 min in AJA International Inc. magnetron sputtering system. Using electron-beam lithography, the transmission line, readout resonators and the gate lines are patterned and etched with ICP-RIE chlorine etching. Using electron beam lithography, fine markers with Ti/Au are defined between the capacitor pads as guides for further steps. Sn/InAs nanowires are positioned within these fine markers using a micro-manipulator. The nanowires are then imaged using SEM to locate the break in the Sn shell that defines the Josephson element. Contacts to the selected nanowires and electrostatic gates are designed using KLayout, a GDS mask layout editor, and defined using electron beam lithography. For Device A and B, Aluminum contacts are deposited using e-beam evaporatore after removing 3~nm of the capping $\text{AlO}_\text{x}$ layer with 80 seconds of 250V~-~15mA in-situ Argon milling. For Devices E and F, NbTiN contacts were sputtered at 23°C. Prior to deposition, the 3 nm $AlO_x$ capping layer was removed via in-situ Argon plasma milling for 40 s (50 W power, 3 mTorr pressure, and 50 sccm Ar flow). Lift-off is performed in acetone and IPA. The devices are then diced and loaded in microwave package using GE varnish.

All measurements discussed in this section are performed in an Oxford triton (DR200) dilution refrigerator. For single tone and two tone spectroscopy measurements, a Keysight P9374A Vector Network Analyzer (VNA) and a SignalCore SC5511A microwave signal generator are used. 

\section{HFSS simulations}

\begin{figure}
    \centering
    \includegraphics[width=0.5\linewidth]{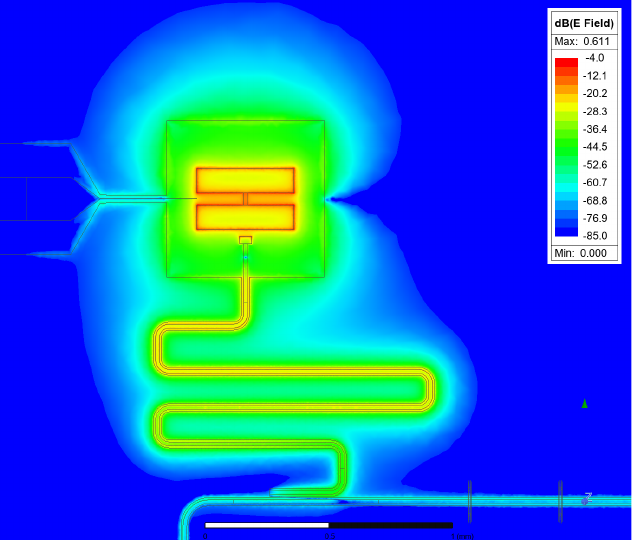}
    \caption{Eigenmode simulation of a single Transmon qubit}
    \label{fig:HFSS-Efields}
\end{figure}

\begin{center}
\begin{table}
\begin{tabularx}{0.8\textwidth} { 
  | >{\centering\arraybackslash}X 
  | >{\centering\arraybackslash}X 
  | >{\centering\arraybackslash}X 
  | >{\centering\arraybackslash}X 
  | >{\centering\arraybackslash}X 
  | >{\centering\arraybackslash}X
  | >{\centering\arraybackslash}X| }
\hline
$f_q$ & $f_r$ &  $E_C/2\pi$ & $\kappa/2\pi$ & $g/2\pi$ & $\chi/2\pi$ & $T_1^{Purcell}$ \\% 1/2pi factors?
\hline
6.67 GHz & 8 GHz & 383.79 MHz & 2-2.5 MHz & -79.47 MHz & -2.43 MHz & 66 us\\
\hline
\end{tabularx}
\caption{Transmon device parameters obtained from Lumped Oscillator Model\cite{Minev2021cQED} simulations}
\label{table:1}
\end{table}
\end{center}

\section{Alternative Explanations: devices E and F.} 

For these two qubits, we used the two highest peaks to extract apparent anharmonicty, assigning those peaks as $f_{01}$ and $f_{02}/2$. We point out that additional lines are visible in the data for qubits E and F. For qubits A and B used in the main text the additional lines are described by transitions to or between higher transmon levels (see main text). The exact origin of additional transitions for qubits D and E is not established, they can also correspond to higher photon transitions, or to artefacts due to measurement circuit nonlinearities. Additional lines can be due to dressed states in the strong coupling regime, however we do not believe this applies to qubits E and F. Due to this, it is possible that the two primary transitions are misidentified for qubits E and F and the anharmonicity should be calculated using farther spaced lines, leading to an overall increase in $\alpha$. In this case all values of $\alpha$ may fall above the minimum short junction limit line. The sensitivity of the frequency difference between the two lines to gate voltage supports the conclusion that these transitions originate from the physics within the nanowire junction.

\section{Additional Data}

\begin{figure*}[!ht]
  \includegraphics[width=14 cm]{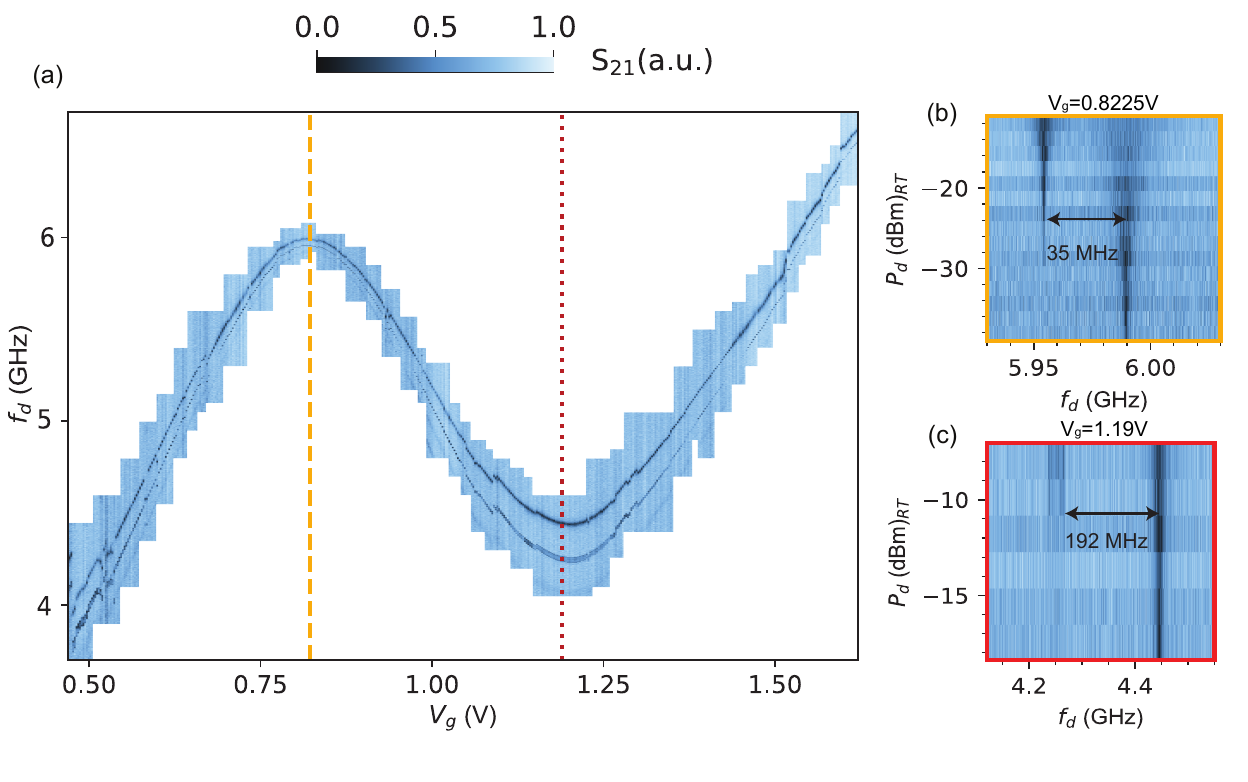}
  \caption{(a) Two-tone spectroscopy as a function of gate voltage $V_g$ and drive frequency $f_d$. (b) Power variation at $V_g=\text{0.8225 V}$ and (c  $V_g=\text{1.19 V}$ showing resolved $f_{01}$ and $f_{02}/2$ at high power.}
  \label{S1-2}
\end{figure*}

\begin{figure*}[!ht]
  \includegraphics[width=\textwidth]{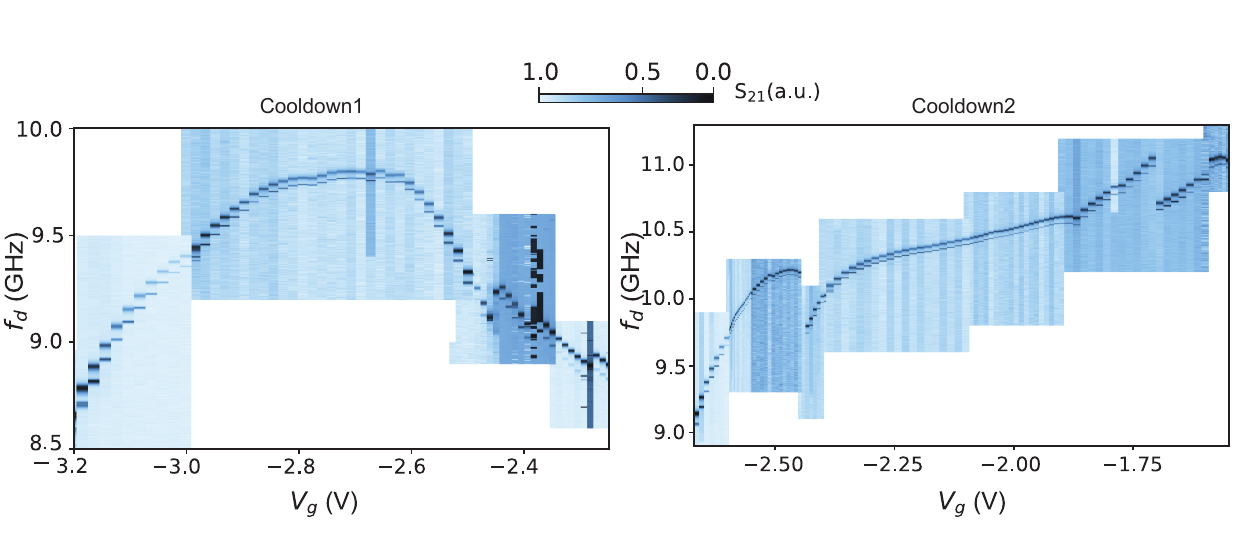}
  \caption{Two-tone spectroscopy as a function of $V_g$ at high drive power for Device E.}
  \label{fig:S1-3}
\end{figure*}

\begin{figure}[!ht]
    \centering    \includegraphics[width=0.5\linewidth]{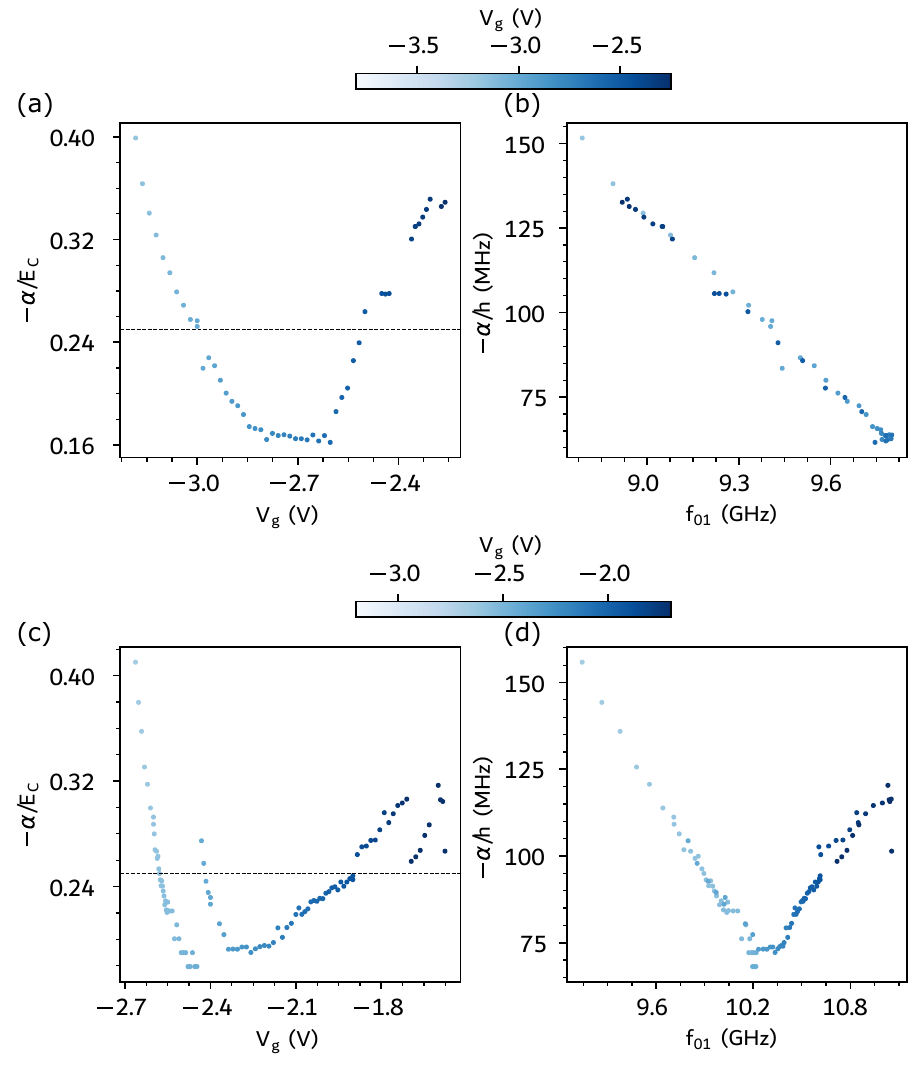}
    \caption{(a) Anharmonicity normalized with the designed charging energy, $\alpha/E_c$ plotted as a function of gate voltage $V_g$, for different qubit frequencies $f_{01}$ indicated by the color bar.(b)Anharmonicity $\alpha/h$ as a function of $f_{01}$, with corresponding gate voltages indicated by the color bar, for Device E, for the first cooldown.(c) and (d) show the same plots as in (a) and (b), respectively, for second cooldown.}
    \label{fig:S1-4}
\end{figure}

\begin{figure}[!ht]
    \centering
    \includegraphics[width=\linewidth]{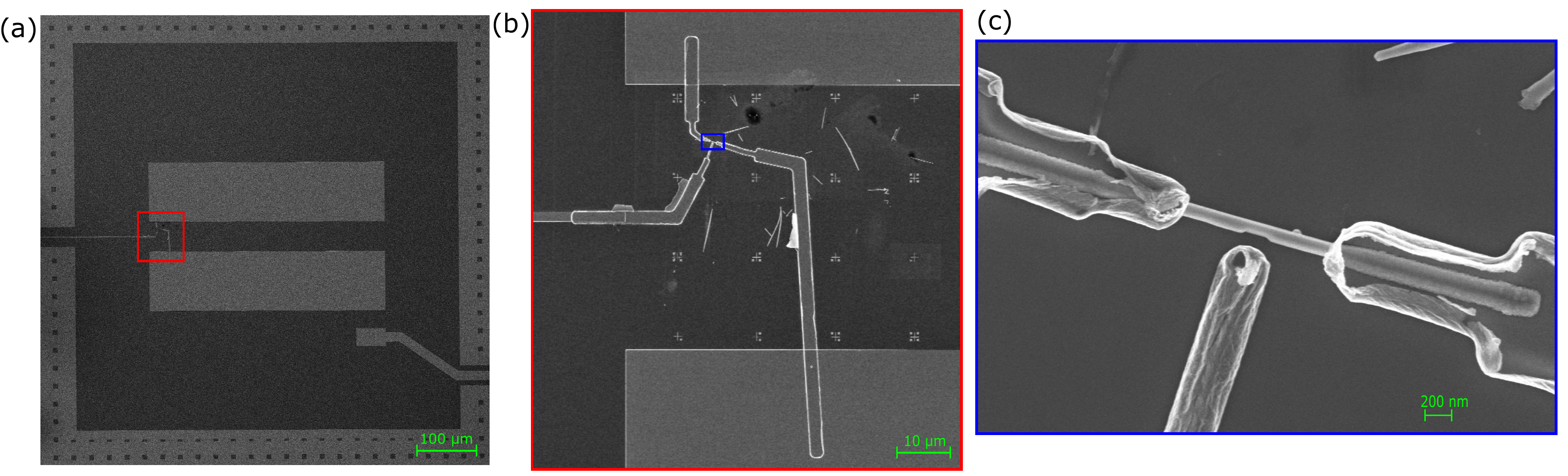}
    \caption{SEM image of Device E}
    \label{fig:S1-5}
\end{figure}

\begin{figure*}[!ht]
  \includegraphics[width=\textwidth]{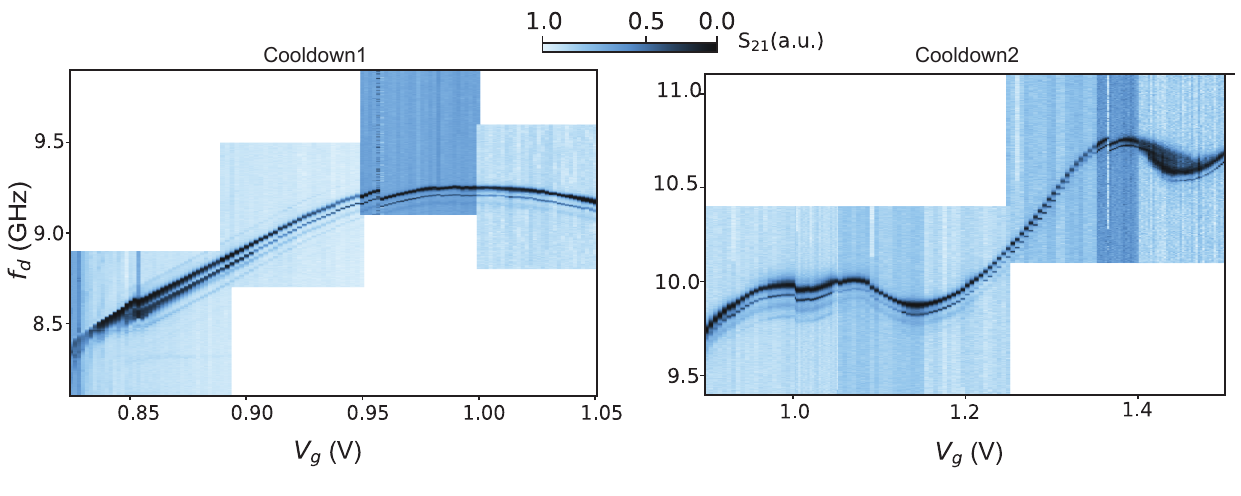}
  \caption{Two-tone spectroscopy as a function of $V_g$ at high drive power for Device F.}
  \label{fig:S1-6}
\end{figure*}

\begin{figure}[!ht]
    \centering
    \includegraphics[width=0.5\linewidth]{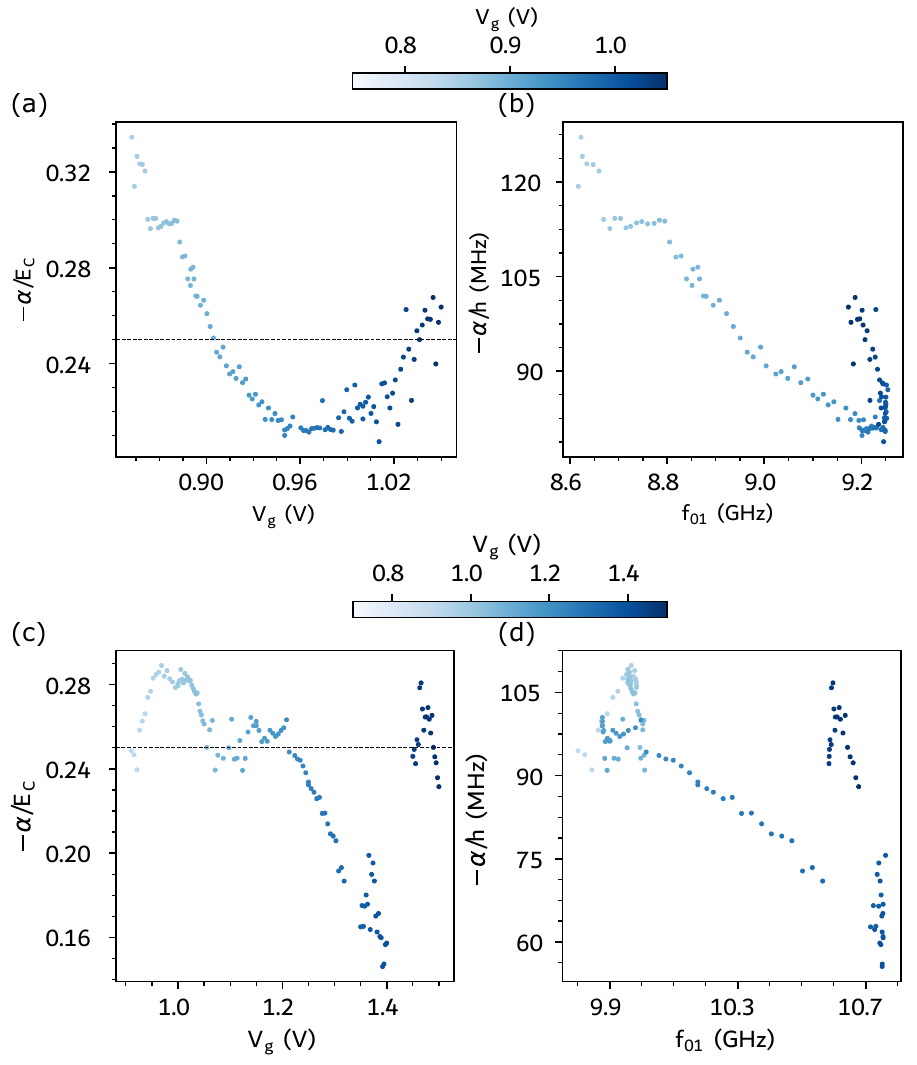}
    \caption{(a) Anharmonicity normalized with the designed charging energy, $\alpha/E_c$ plotted as a function of gate voltage $V_g$, for different qubit frequencies $f_{01}$ indicated by the color bar.(b)Anharmonicity $\alpha/h$ as a function of $f_{01}$, with corresponding gate voltages indicated by the color bar, for Device F, for the first cooldown.(c) and (d) show the same plots as in (a) and (b), respectively, for second cooldown.}
    \label{fig:S1-7}
\end{figure}

\begin{figure}[!ht]
    \centering
    \includegraphics[width=\linewidth]{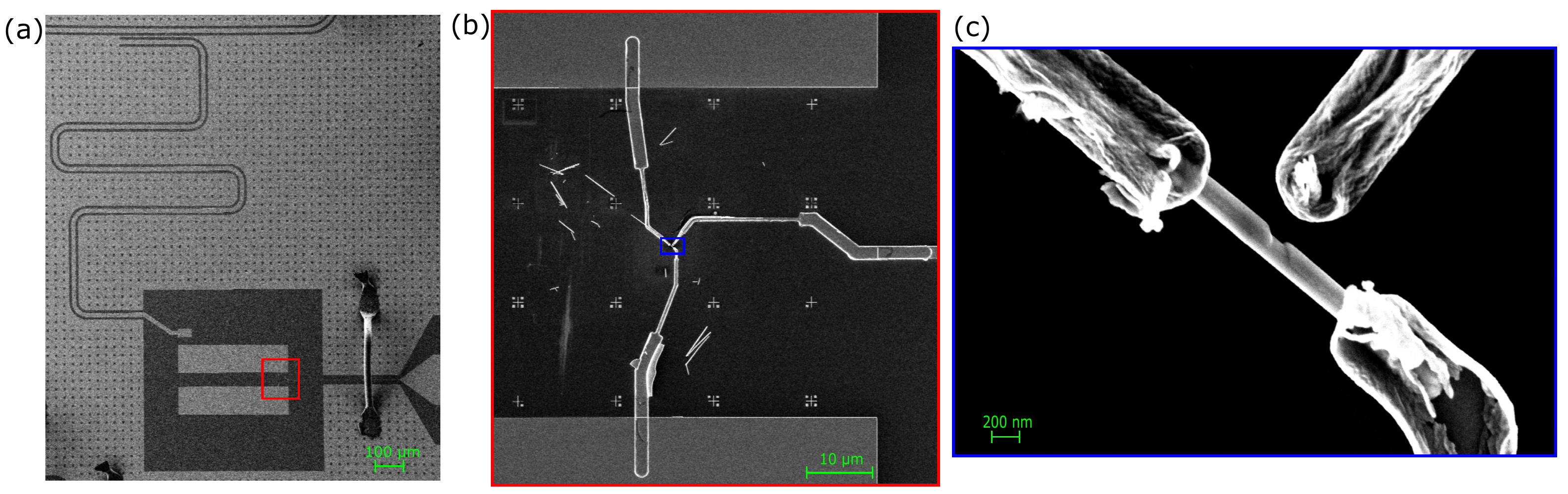}
    \caption{SEM image of Device F}
    \label{fig:S1-8}
\end{figure}

\end{document}